\documentclass{article}
\usepackage{graphicx} 
\usepackage{booktabs}  
\usepackage[backend=bibtex,style=numeric,sorting=none]{biblatex}
\usepackage{hyperref}
\usepackage{multirow}
\usepackage{amsmath}
\usepackage{longtable}  
\usepackage{array} 
\usepackage{changepage} 
\usepackage[margin=1in]{geometry}
\usepackage[symbol]{footmisc}
\hypersetup{
    colorlinks=true,        
    linkcolor=black,        
    citecolor=blue,         
    urlcolor=blue,          
    pdfborder={0 0 0},      
    hyperfootnotes=false    
}

\newcommand{\imputed}{\textsuperscript{\dag}}
\title{Measuring AI Diffusion: A Population-Normalized Metric for Tracking Global AI Usage}
\author{Amit Misra\thanks{These authors contributed equally to this work.}, Jane Wang\footnotemark[1], Scott McCullers, Kevin White, and Juan Lavista Ferres \\ \\
Microsoft AI for Good Lab
}

\date{}

\begin{document}
\maketitle
\setcounter{footnote}{1}
\begin{abstract}
    Measuring global AI diffusion remains challenging due to a lack of population-normalized, cross-country usage data. We introduce AI User Share, a novel indicator that estimates the share of each country's working-age population actively using AI tools. Built from anonymized Microsoft telemetry and adjusted for device access and mobile scaling, this metric spans 147 economies and provides consistent, real-time insight into global AI diffusion. We find wide variation in adoption, with a strong correlation between AI User Share and GDP. High uptake is concentrated in developed economies, though usage among internet-connected populations in lower-income countries reveals substantial latent demand. We also detect sharp increases in usage following major product launches, such as DeepSeek in early 2025. While the metric's reliance solely on Microsoft telemetry introduces potential biases related to this user base, it offers an important new lens into how AI is spreading globally. AI User Share enables timely benchmarking that can inform data-driven AI policy. 
\end{abstract}

\section{Introduction}

Artificial intelligence (AI) has rapidly ascended from a niche technology to a globally pervasive general-purpose tool, with adoption accelerating dramatically in recent years. Worldwide, the share of organizations implementing AI reached about 78\% in 2024, up from 55\% the year before~\supercite{maslej_artificial_2025}. On the consumer side, the advent of generative AI has driven an unprecedented wave of usage. For example, OpenAI’s ChatGPT surpassed 1 million users in just 5 days after launching in November 2022, and reached 400 million users by February 2025~\supercite{noauthor_number_2024}—a pace of uptake far exceeding earlier general purpose technologies such as the internet or mobile phones~\supercite{bick2024rapid}. This meteoric rise underscores AI’s broad appeal and transformative potential across economies. 

However, the diffusion of AI is unfolding unevenly, with stark regional and income-level disparities. High-income countries currently lead in AI utilization, while many lower-income nations remain on the periphery of this digital revolution~\supercite{khan_artificial_2024}. These disparities reflect longstanding digital divides: Wealthier economies benefit from superior digital infrastructure, affordability, and human capital—factors strongly predictive of higher AI uptake~\supercite{liu_who_2024}—whereas developing economies face greater barriers to access. Notably, only 27\% of individuals in low-income countries have reliable internet access~\supercite{ITU_FactsFigures_2024}, highlighting the foundational gap underlying differences in advanced technology use. 

The dynamic nature of the AI market further shapes global usage patterns. Technological breakthroughs and competitive forces can swiftly alter the trajectory of adoption. A prominent recent example is the launch of DeepSeek in January 2025, which rapidly accelerated global AI uptake. Within days of its debut, DeepSeek’s AI assistant overtook ChatGPT as the top-rated free application on the U.S. Apple App Store~\supercite{baptista_what_2025}, demonstrating a surge in new users attracted by its accessibility and low cost. 

Despite the growing importance of AI diffusion, there is a significant lack of consistent, population-wide data on actual AI usage at a country\footnote{In this paper, the term “Country”—also referred to as “Economy”—encompasses any distinct country, territory, or jurisdiction for which independent economic and social data are compiled and reported.
Our use of Country/Economy reflects data‐availability and reporting conventions established by organizations such as the World Bank, the International Monetary Fund, and the United Nations. It does not imply recognition of political sovereignty or legal status; rather, it ensures consistency and clarity when comparing metrics across diverse geopolitical entities.} level. For instance, the World Bank’s recent report employs Semrush traffic data to estimate generative AI adoption among internet users~\supercite{liu_who_2024}, yet this approach overlooks approximately one-third of the global population lacking reliable internet access~\supercite{noauthor_world_nodate}. Most country-level assessments instead rely on ad hoc proxies, surveys, or firm-level metrics rather than comprehensive quantitative usage data. The Pew Research Center reports that 23\% of U.S. adults have used ChatGPT~\supercite{mcclain_americans_2024}, yet comparable surveys are scarce outside the United States. Similarly, industry studies such as McKinsey’s 2025 Global AI Survey emphasize organizational adoption, noting that 78\% of companies use AI in at least one business function, but fail to measure individual-level adoption across countries~\supercite{mckinsey_state_of_ai}. Khan et al. explicitly acknowledge that “the limited presence of AI in low-income countries prevents quantitative investigation due to the lack of available data on national AI adoption and usage”~\supercite{khan_artificial_2024}. Thus, no consistent metric currently captures the share of the entire working-age population using AI across countries. 

To address this gap, we develop a global, population-normalized metric— AI User Share — derived from Microsoft telemetry. By combining (1) the share of users engaging with AI, (2) desktop device penetration estimates, and (3) a country-specific mobile-to-desktop usage ratio, we estimate the number of individuals actively using AI in each country. This approach provides country-level resolution in tracking global adoption trends. 

In this paper, we first detail our methodology for calculating the AI User Share metric, and then highlight key insights from our analysis. Specifically, we present (1) comparative AI diffusion metrics across 147 countries and regions, (2) correlations between AI usage and economic strength, (3) the impact of internet access on AI User share and (4) the temporal impact of DeepSeek’s launch on global AI adoption trends. Our findings provide a unified, data-driven perspective on global AI diffusion, equipping policymakers and researchers with a consistent benchmark to monitor and foster AI adoption worldwide. 

\section{Methods}

We use Microsoft-proprietary usage data to estimate AI usage across key AI services. This data provides a rich view into user behaviors across the internet and allows us to estimate the number of AI users over time by country. Our overall methodology is to estimate the AI User Share per country per time period as follows:

\begin{align}  
\text{AI User Share} = & \left( \%\ \text{of Microsoft Users That Use AI} \right) \notag \\  
& \times \left( \%\ \text{of Population With a Desktop Device} \right) \notag \\  
& \times \left(\text{Mobile Scaling Factor} \right)
\end{align}  

To get the absolute number of AI users we multiply the AI User Share by the working-age population. The precise calculation involves additional steps detailed below, including accounting for outliers and adjusting for potential overlap between desktop and mobile usage.

\subsection{Percent of Microsoft Users that Use AI}

Our estimation of AI usage begins with anonymized telemetry data from Microsoft, primarily originating from users on Windows desktop (PC/tablet) platforms. This dataset is subject to user privacy controls, allowing individuals to opt out of sending diagnostic data, which impacts overall data completeness. Despite these factors, the logs enable the identification of visits to specific AI sites (e.g., ChatGPT, Gemini, Claude, Microsoft Copilot; see Appendix for full list) with high precision across many global regions. The following steps detail our approach to filtering for active usage and adjusting for potential biases introduced by infrequent users and telemetry opt-out rates.

One challenge with our data is that it includes many infrequent users who only use the product a few times during the month. This can be due to a number of factors, including low internet usage for the user overall, or that the majority of their usage happening on another device or in another product. To filter out these users, we only consider users with at least 90 minutes of usage time in a given month (3 minutes a day) in our analysis. While this amount of usage is fairly low, this threshold is meant to filter out the infrequent users who could skew our AI usage numbers downward. 

Another challenge with using internal Microsoft data is that not all users choose to share telemetry with us. Many users opt out of sending diagnostic data, meaning we do not have a complete view of all user activity.

In many countries, the majority of users do opt in, providing us with a reasonably representative sample. While this subset of users may not perfectly reflect the population as a whole, we are confident that the data captures meaningful patterns in AI usage.

However, for countries where the vast majority of our users opt out of sending telemetry to us - particularly in parts of Europe - our sample may be too limited to reliably reflect local trends. To account for this, we adjust AI usage estimates in markets with below-average opt-in rates by blending the observed country-level usage with the global average. The lower the opt-in rate, the more heavily we weight the global average. This approach helps stabilize estimates in data-sparse regions and reduces the risk of over-interpreting noisy or unrepresentative samples.

Specifically, we apply the following adjustment:
\[  
\begin{cases}   
\gamma, & \text{if } \alpha > \bar{\alpha} \\   
\bar{\gamma} \cdot \left(\frac{\bar{\alpha} - \alpha}{\bar{\alpha}}\right) + \gamma \cdot \frac{\alpha}{\bar{\alpha}}, & \text{otherwise}  
\end{cases}  
\]  
  
\noindent where:
\begin{itemize}
    \item \( \alpha \) is the country-level opt-in rate
    \item \( \bar{\alpha} \) is the global average opt-in rate
    \item \( \gamma \) is the raw AI usage share for the country
    \item \( \bar{\gamma} \) is the global average AI usage share
\end{itemize}

\noindent When a country’s opt-in rate exceeds the global average, we use the raw value under the assumption that the data is sufficiently representative.

\subsection{Adjusting for Desktop Device Penetration and Mobile Usage}

A key challenge in extrapolating from internal Microsoft data to the broader population is that not all individuals have access to a device. To adjust for this, we use two scale factors to account for Desktop device penetration and for mobile usage. We use two different scalings to reflect that our data is primarily desktop-focused. Therefore, we first extrapolate from Microsoft desktop data to overall desktop usage, then account for mobile usage. 

We estimate the proportion of the working-age population in each country with a PC or tablet using a combination of Microsoft internal telemetry and third-party data. We begin with Windows telemetry, which provides a count of monthly active devices (MAD) in each country. We then divide this figure by the Windows market share for PC and tablet devices — sourced from StatCounter~\supercite{noauthor_desktop_nodate-1} — to estimate the total number of such devices across all platforms.

Next, we divide the total estimated device count by the country’s working-age population to obtain a per-capita device access ratio. Since this ratio can exceed 1 in high-access countries, we normalize the values using the following:

\begin{align}  
\text{Numerator} = & \ \text{Device Ratio}_{\text{country}} - \min\left(\text{Device Ratio}\right) \notag \\
\text{Denominator} = & \ \text{Percentile}_{90}\left(\text{Device Ratio}\right) - \min\left(\text{Device Ratio}\right) \notag \\
\text{Scaling Factor} = & \ \left( \frac{\text{Numerator}}{\text{Denominator}} \right) \times 0.9 + 0.1
\end{align}

Here, \textit{Device Ratio} refers to the number of PC and tablet devices per working-age person in a given country. The 90th percentile cap prevents extremely high-access countries from disproportionately skewing the normalization.

The additive constant of 0.1 ensures that all resulting scaling factors remain greater than zero and retain meaningful interpretability, even in countries with very limited technology access. Without this adjustment, low-access countries could yield unrealistically low AI usage estimates.


The methodology described above is focused on AI usage on desktop devices. However, we know that a significant share of AI usage occurs on mobile. To account for this, we apply a country-specific mobile scaling factor to extrapolate our desktop-based estimates to include mobile usage.

We derive this scaling factor using third-party data from StatCounter~\supercite{noauthor_desktop_nodate}, which provides estimates of the mobile-to-desktop traffic ratio for each country. In cases where a country's ratio is more than 1.8 times higher than the regional average, we substitute the regional average instead. The 1.8× threshold was derived from the global distribution of desktop usage by country, and corresponds to a country being more than 1.28 standard deviations above the global mean (roughly the 90th percentile).

Finally, to estimate the total AI User Share across both platforms we adjust for potential overlapping users. Assuming statistical independence\footnote{This assumption of statistical independence is a methodological simplification. Actual overlap between desktop and mobile AI usage is likely positively correlated, meaning the true overlap probably exceeds the estimate derived under independence (the product of the shares). This could result in a slight overestimation of the final combined AI User Share.} between desktop and mobile AI usage patterns, we estimate the overlap share by multiplying the estimated desktop AI user share (derived above using Microsoft data and desktop device penetration) by the estimated mobile AI user share (which we derive by applying the mobile scaling factor to the desktop share). This product approximates the share of dual-platform users. We then subtract this estimated overlap from the simple sum of the desktop and mobile shares to arrive at the final AI User Share, correcting for double-counting.

The extrapolation from our desktop estimates to combined desktop and mobile numbers rests on multiple assumptions. First, we assume that the mobile-to-desktop ratio for AI usage mirrors the overall internet usage ratio. Second, we assume that mobile and desktop usage behaviors are sufficiently similar, and that this relationship remains stable over time.

While these assumptions are difficult to validate, we note that this scaling does not impact the underlying desktop trends we estimate from Microsoft data; it simply provides a more comprehensive view of total AI usage across platforms.

\subsection{Working-Age Population}

To estimate the total number of AI users, we multiply the AI Usage Share by the working-age (15-64 years) population in each country using data from the World Bank's World Development Indicators~\supercite{noauthor_world_nodate}. For countries where World Bank lacks data, we use  data from CEIC Data~\supercite{noauthor_global_2025}. We focus on the working-age population rather than the total population for several reasons.

First, this age range is the standard international benchmark adopted by organizations like the World Bank, International Labour Organization (ILO)~\supercite{international_labour_organization_world_2023} and Organization for Economic Co-operation and Development (OECD), primarily to ensure consistent cross-country data comparability~\supercite{noauthor_working_nodate}. It serves as a widely available proxy for the population segment most likely to be economically active. While related metrics, such as internet usage definitions cited by the ITU often use slightly different age criteria (e.g., 15+), employing the standard 15-64 definition for our population denominator grounds our AI penetration measure in established global demographic and economic benchmarks, facilitating broader comparisons. Additionally, analysis from the Pew Research Center shows that only 6\% of US adults age 65 and up have tried ChatGPT~\supercite{mcclain_americans_2024}, suggesting that restricting to the 15-64 year old age range captures the vast majority of AI usage.


Second, our data is derived from PC and tablet usage, which are typically used in work or productivity settings. These devices are less common among children and older adults, who are also less likely to use AI tools regularly.

Finally, using the working-age population as a denominator avoids diluting the AI user share with demographic groups that are less relevant to this analysis, and results in a more meaningful measure of adoption within the active digital workforce.

\subsection{Data Scope and Filtering}

To ensure representative coverage across regions and account for potential seasonal effects, we aggregate across data going back to late 2024 in all analyses unless otherwise indicated. This extended window improves robustness in cross-country comparisons and mitigates short-term fluctuations.

Additionally, to ensure the reliability of our findings, we restricted any country-specific analyses to countries with sufficient volume of monthly traffic to generate robust estimates and a minimum total population of 2 million. Countries with limited traffic data or very small populations were grouped into small regions when possible, and the data from all countries in a region were aggregated to calculate a regional AI usage share metric. For example, we created an East Africa region that includes Burundi, Eritrea, Ethiopia, Somalia, South Sudan, Sudan, Tanzania, and Uganda - all countries that fell below our thresholds and therefore needed to be grouped with other countries.

\subsection*{Data Coverage Disclaimer}
AI User Share estimates are based on Microsoft telemetry data. In some countries—like Russia, Iran, and partly China—telemetry data is limited, so usage estimates may be incomplete. These gaps should be kept in mind when looking at country-level AI adoption.

\section{Results}

We report on the current state of global AI adoption using the AI User Share metric, based on Microsoft telemetry and scaled to the working-age population. Our findings highlight significant geographic disparities, a strong correlation with national wealth, and the impact of product launches and internet access. Results are presented in four parts: global variation, economic correlations, connected user insights, and recent product impacts.

\subsection{Global Variation in AI Adoption Rates}

\begin{figure}[ht]
    \centering
    \includegraphics[width=\textwidth]{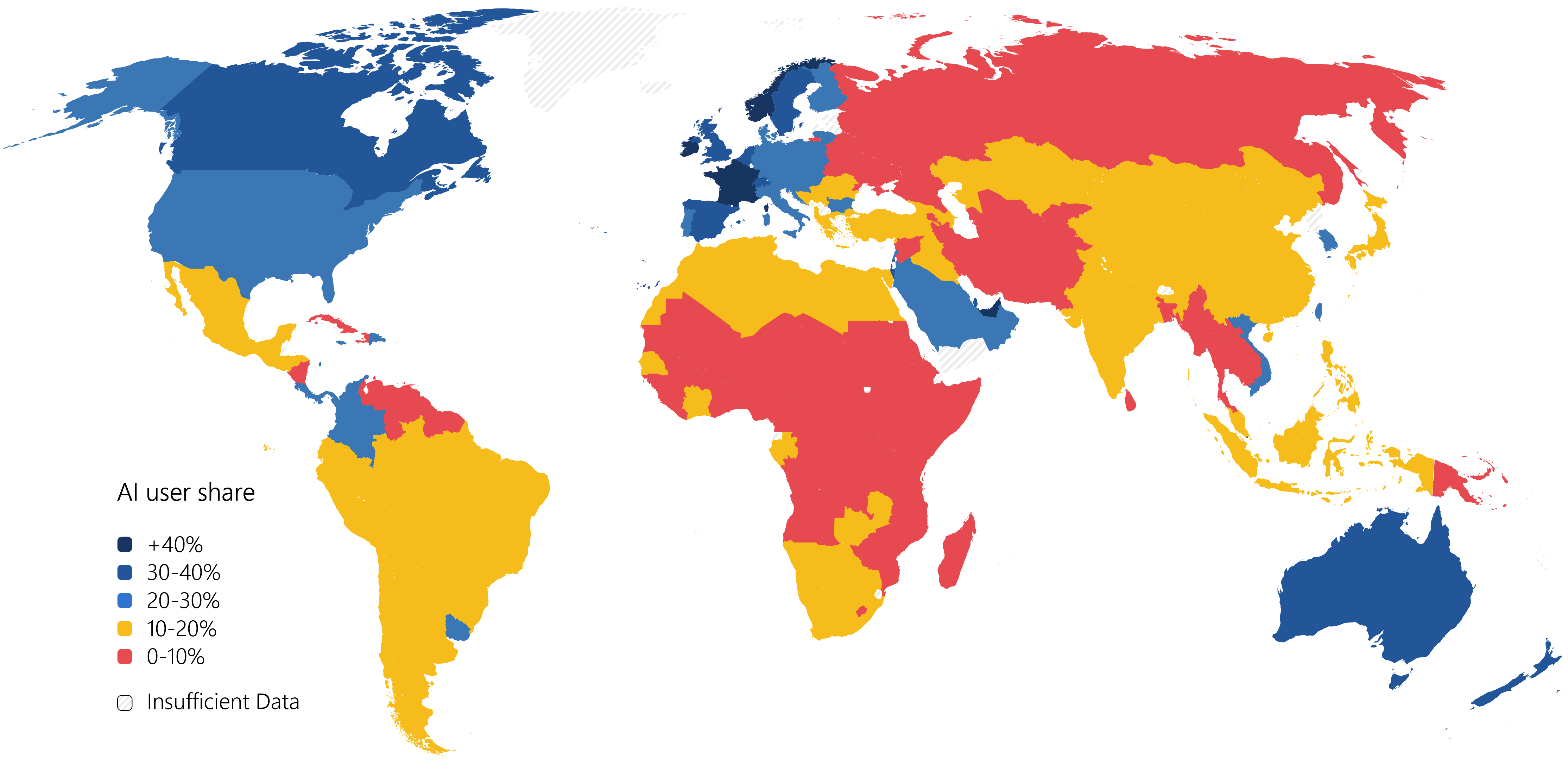}
    \caption{
        Global distribution of AI adoption rates by economy. The United Arab Emirates and Singapore exhibit the highest adoption, with over half of the working-age population using AI tools. Most high-adoption economies are in Europe and North America.
    }
    \label{fig:ai_adoption_map}
\end{figure}

Levels of AI adoption differ significantly by country/region, with a global average of 15\% of the working-age population using AI. UAE and Singapore lead globally, each with an AI User Share of 59\% Figure \ref{fig:ai_adoption_map} shows the global view of AI Adoption rates by economy\footnote{The world map included in this report is provided solely for illustrative purposes. The boundaries and territorial delineations shown do not imply any endorsement, recognition, or opinion by the authors, their institutions, or data providers regarding political sovereignty, legal status, or territorial disputes.}. AI adoption levels vary significantly across countries and regions. North America leads with 27\% of its working-age population using AI tools, followed by Europe and Central Asia at 22\%. Conversely, lower adoption rates are observed in South Asia and Sub-Saharan Africa, suggesting a persistent global digital divide in AI usage. European countries dominate the top 30, accounting for 18 entries (see Table \ref{tab:ai_users_top30}), indicating a concentration of adoption among developed economies.

\begin{figure}[ht]
    \centering
    \includegraphics[width=\textwidth]{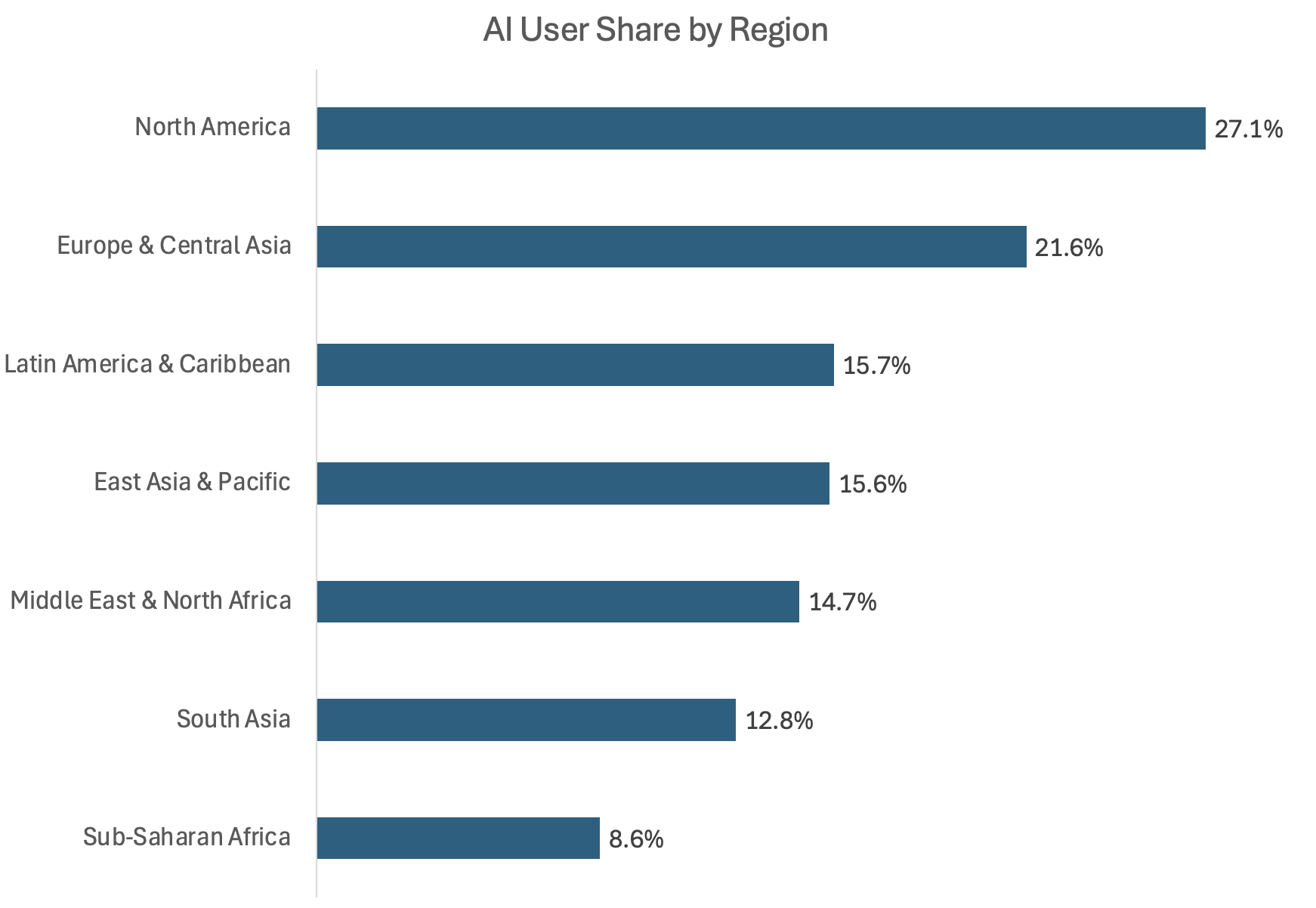}
    \caption{Regional comparison of AI user share as of June 2025. North America and Europe and Central Asia exhibit the highest adoption rates, while South Asia and Sub-Saharan Africa remain below 13\%.}
    \label{fig:ai_adoption_regions}
\end{figure}

\begin{table}[ht]
\centering
\caption{Top 30 Economies by AI User Share}
\label{tab:ai_users_top30}
\begin{tabular}{|l|c|c|}
\hline
\textbf{Rank} & \textbf{Economy} & \textbf{AI User Share} \\
\hline
1 & United Arab Emirates & 59.4\%\\
2 & Singapore & 58.6\%\\
3 & Norway & 45.3\%\\
4 & Ireland & 41.7\%\\
5 & France & 40.9\%\\
6 & Spain & 39.7\%\\
7 & New Zealand & 37.6\%\\
8 & United Kingdom & 36.4\%\\
9 & Netherlands & 36.3\%\\
10 & Qatar & 35.7\%\\
11 & Australia & 34.5\%\\
12 & Israel & 33.9\%\\
13 & Canada & 33.5\%\\
14 & Belgium & 33.5\%\\
15 & Switzerland & 32.4\%\\
16 & Sweden & 31.2\%\\
17 & Austria & 29.1\%\\
18 & Hungary & 27.9\%\\
19 & Denmark & 26.6\%\\
20 & Germany & 26.5\%\\
21 & Poland & 26.4\%\\
22 & Taiwan & 26.4\%\\
23 & United States & 26.3\%\\
24 & Czech Republic & 26.0\%\\
25 & South Korea & 25.9\%\\
26 & Italy & 25.8\%\\
27 & Finland & 25.6\%\\
28 & Bulgaria & 25.4\%\\
29 & Jordan & 25.4\%\\
30 & Costa Rica & 25.1\%\\
\hline
\end{tabular}
\end{table}

\clearpage

\subsection{Correlation Between Economic Strength and AI Adoption}

\begin{figure}[ht]
    \centering
    \includegraphics[width=\textwidth]{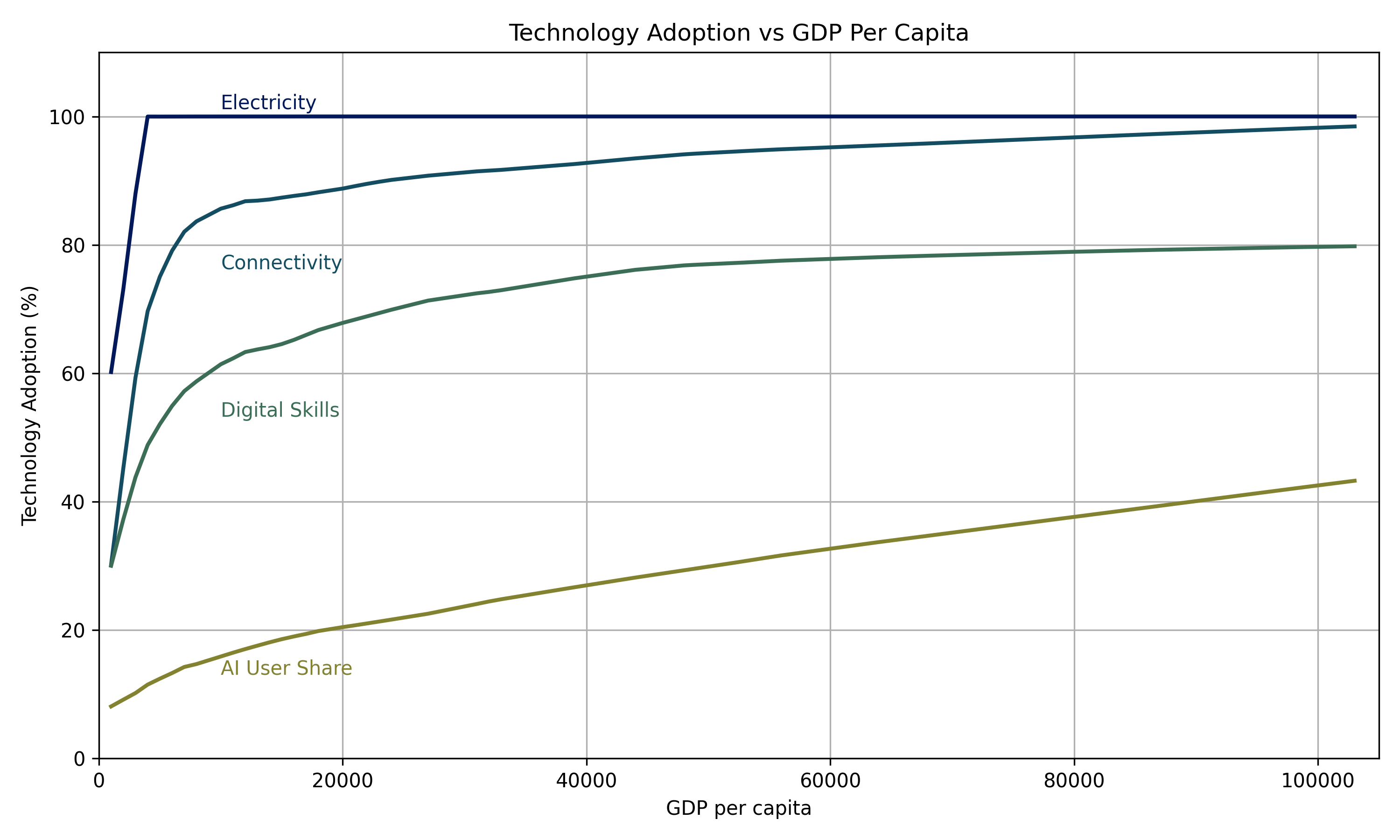}
    \caption{Technology adoption vs GDP per Capita for electricity, internet access, digital skills and AI User Share. All four technologies are highly correlated with GDP per Capita, though the absolute level of adoption decreases as we progress from electricity access to AI User Share.}
    \label{fig:tech_adoption}
\end{figure}

To better understand the link between economic strength and AI adoption, we examined how GDP per capita relates to the share of AI users across countries. First, we look at aggregate trends for how different technologies correlate with GDP per capita~\supercite{noauthor_world_nodate}, as shown in Figure \ref{fig:tech_adoption}. We examine electricity~\supercite{IEA_IRENA_UNSD_WorldBank_WHO_2023}, internet connectivity~\supercite{ITU_DataHub_11624, ITU_FactsFigures_2024}, the World Bank Digital Adoption Index~\supercite{WB_DAI, WDR2016} as a proxy for digital skills, and AI User Share. All of these technologies are highly correlated with GDP per capita. These four can be considered a funnel, since it is difficult to use AI without having basic digital skills. Likewise, basic digital skills are unlikely to be present without internet connectivity, and even moreso, electricity. Figure \ref{fig:tech_adoption} shows the dropoff in each stage of the funnel as a function of GDP per capita, highlighting the gap between AI usage and other technologies - in other words, AI usage has a considerable way to go before becoming as prevalent as digital skills.

\begin{figure}[ht]
    \centering
    \includegraphics[width=\textwidth]{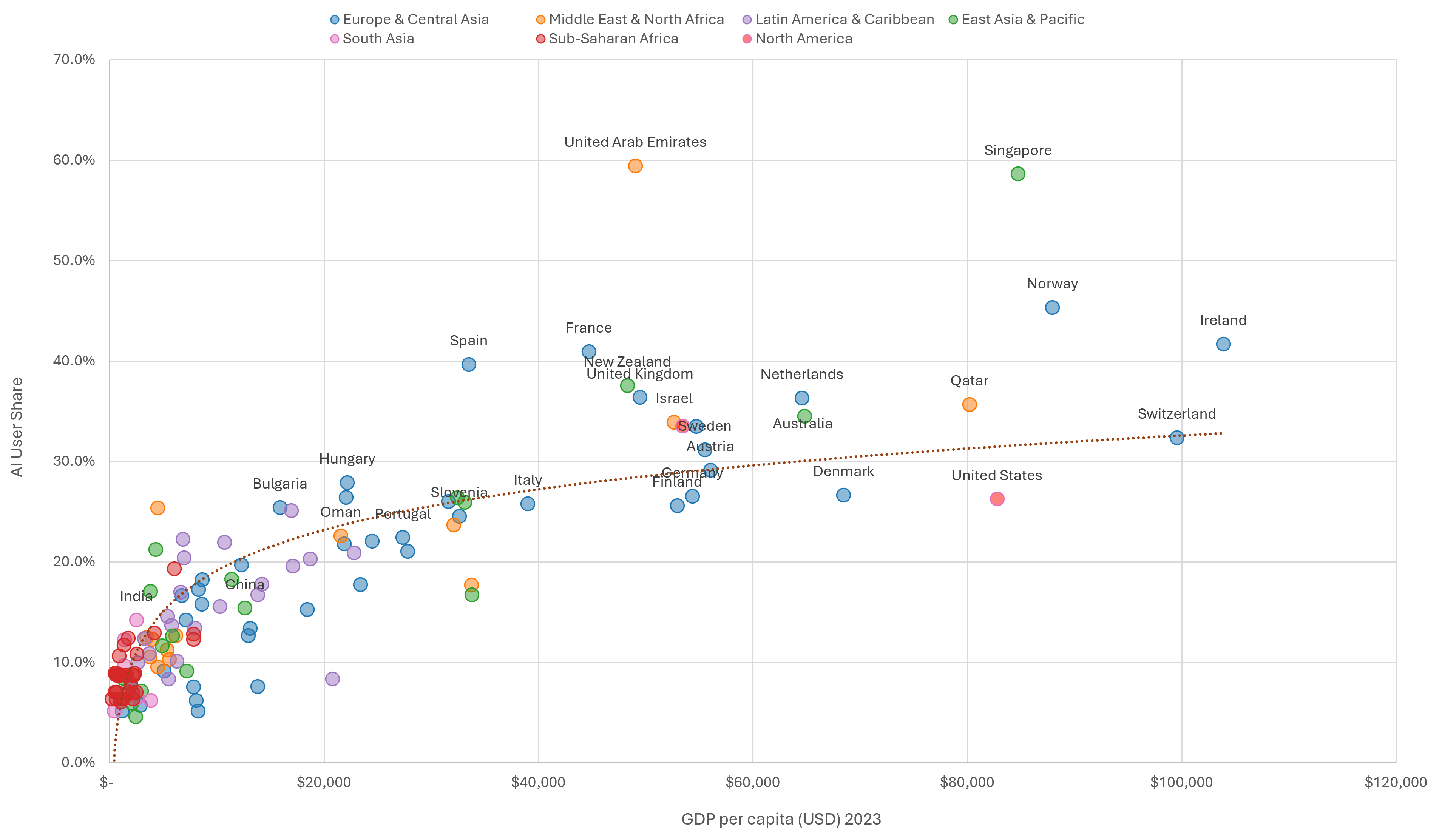}
    \caption{Comparison of AI Adoption Rates with GDP Per Capita. In general, there is an upward trend with higher AI adoption as GDP per capita increases. However, there appears to be a leveling off in the range of 25-40\%, implying a potential ceiling to AI adoption given current conditions.}
    \label{fig:ai_vs_gdp}
\end{figure}

Exploring more deeply the relationship between AI usage and economic strength, Figure \ref{fig:ai_vs_gdp} shows the relationship between AI Adoption Rates and GDP per capita, along with a log-linear trend. The analysis produced a Spearman correlation coefficient of 0.83, with a p-value below 0.000001, indicating a strong statistically significant positive correlation. This suggests AI user share is closely linked to a country’s economic strength, though growth appears more gradual at higher income levels. Aside from outliers like UAE and Singapore , most high-income countries cluster within the 25-45\% range. 

Additionally, a few advanced economies, including the United States and Denmark, fall below the trendline, indicating they are underperforming relative to peers with similar income levels. These gaps suggest opportunities for further growth, it may also reflect differences in public sentiment, education, workplace AI integration, or AI regulatory environment - all of which warrant further research.  

\subsection{AI Adoption Among Connected Populations}

While overall AI adoption remains relatively low in many South Asian, African and Latin American economies, a clearer picture emerges when looking specifically at the population with internet access. We therefore examined AI User Share (Connected Population), defined as the AI User Share of a percent of the population with internet access (based on data from ITU~\supercite{ITU_DataHub_11624}), in a selected group of economies.

\begin{figure}[ht]
    \centering
    \includegraphics[width=\textwidth]{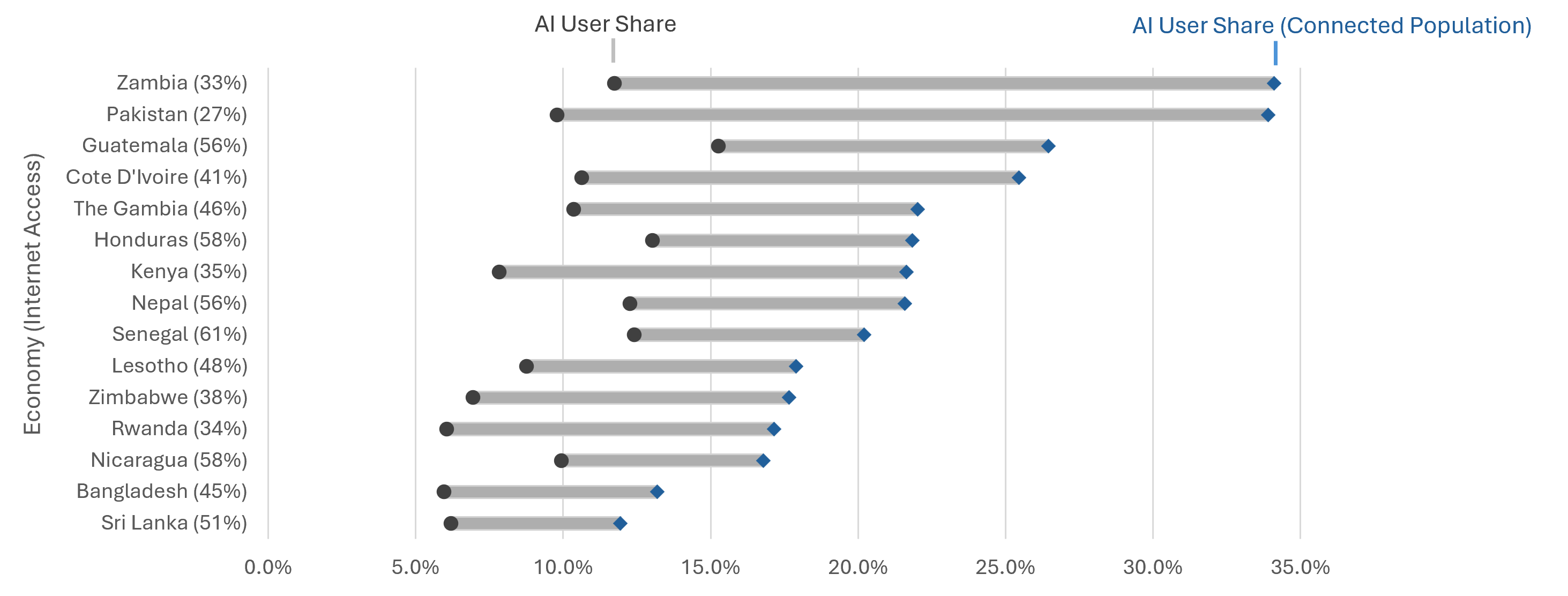}
    \caption{AI User Share as a percent of the connected population for countries with the lowest internet penetration. When normalized to population connected to the internet, many of these countries show high rates of AI adoption.}
    \label{fig:ai_connected_populations}
\end{figure}

To illustrate the opportunity gap, we focused on the 15 economies with the lowest internet penetration among the 111 economies in our study that we are able to estimate economy-specific estimates for (as opposed to regional averages for the economies with insufficient data). The AI User Share (Connected Population) for the 15 economies is shown in Figure \ref{fig:ai_connected_populations}. We see that many countries are already experiencing meaningful levels of engagement among their online populations. In Zambia, Pakistan, Guatemala and Côte d’Ivoire, over one in four connected individuals is already using AI tools. Zambia’s AI User Share rises from 12\% at the population level to 34\% among its connected population, and Pakistan jumps from 10\% to 33\%, demonstrating substantial enthusiasm for AI where internet access exists. 

This pattern of high engagement among the connected holds across the group. Averaged across these 15 low-penetration economies, the AI User Share (Connected Population) reaches 23\%, even higher than the 20\% connected-population average of all other countries with sufficient country-specific data.  In stark contrast, their overall AI User Share averages only 9\%, compared to 17\% for the broader set of 111 countries analyzed. This shift—from trailing by 8 percentage points overall to leading by 3 points among connected users—strongly suggests that the primary barrier to wider AI use in these nations is access, not a lack of demand among those already online.

This contrast highlights a critical opportunity: although many of these countries currently face limited internet access (with coverage ranging from 27\% to 61\%), those who are connected are not just keeping pace but in fact adopting AI at rates that match or exceed global average.

\subsection{The Impact of DeepSeek}

\begin{figure}[ht]
    \centering
    \includegraphics[width=\textwidth]{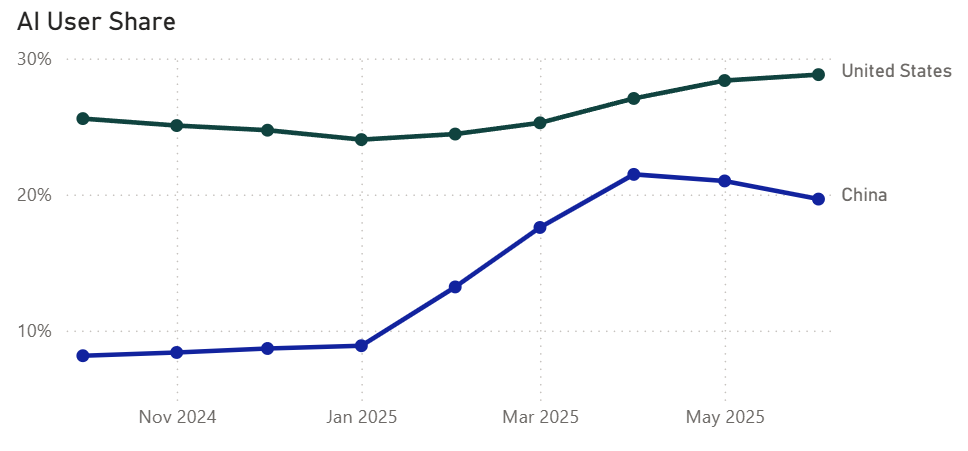}
    \caption{Comparison of China and US AI User Share over time (3-month rolling average). After the launch of DeepSeek's latest model in January 2025, China's AI User Share has more than doubled to 20\%.}
    \label{fig:china_vs_us}
\end{figure}

The launch of DeepSeek in January 2025 resulted in accelerated global AI adoption by driving direct user growth. Figure \ref{fig:china_vs_us} shows the change in AI User Share in the United States and China using a 3-month rolling average. Over this time period, the United States has maintained a steady AI user share of around 25\%. In contrast, China is rapidly closing the gap, driven by increased DeepSeek users. Since then, China’s AI user share has more than doubled from 8\% to 20\%, making it the world’s largest AI market, with an estimated AI user base exceeding 195 million. The growth in China’s AI user population also appears to be sustained, suggesting that DeepSeek is not only attracting new users, but also keeping them engaged.  

\section{Discussion and Conclusions}

This paper introduced the AI User Share, a novel metric quantifying AI adoption as the share of the entire working-age population actively using AI tools globally. Addressing limitations of prior approaches—which often rely on self-reported surveys, firm-level data, or usage metrics confined to internet users—our methodology leverages direct usage from Microsoft telemetry. By integrating the proportion of active Microsoft AI users with estimates of device penetration and mobile usage patterns, we generate a consistent, country-level measure of actual AI engagement across the full demographic. This population-centric framework provides a vital benchmark for policymakers and researchers needing a reliable, globally comparable view to track AI diffusion and inform strategies for equitable adoption. 


In comparing our country and region level rankings with existing AI indices, we find significant alignment in the top-ranked countries and regions. The top 5 countries in our AI Usage share metric (UAE, Singapore, France, Ireland, and Norway) all consistently rank within the top 20 across multiple industry and institutional reports~\supercite{nettel_government_2024, noauthor_global_nodate-1, fattorini_global_nodate}. Similarly, North America and Europe consistently rank highly in both our analyses and other studies, such as the Oxford Government AI Readiness Index~\supercite{nettel_government_2024}). However, many prominent AI indices omit low-income countries entirely. For example, Stanford HAI’s Global AI Vibrancy Tool~\supercite{fattorini_global_nodate} ranks only 36 countries—predominantly high and upper-middle-income economies—leaving out most low and lower-middle-income nations. Likewise, the OECD AI Policy Observatory’s database~\supercite{oecd_ai_dashboard_2021} (OECD's live repository of AI strategies \& policies) covers 69 jurisdictions, including OECD members and select partners, but still excludes many countries where digital access remains limited.  By contrast, our AI User Share metric spans 147 countries and regions with sufficient Microsoft data, uncovering usage in regions that traditional rankings overlook. 

 To further contextualize our results, we compare AI User Share ranking with country-level metrics published by the World Bank~\supercite{liu_who_2024}, which ranks economies based on ChatGPT site traffic per internet user. While we observe general alignment among several high-adoption countries such as Singapore, the United States, and Canada, the differences are notable in smaller economies. The World Bank's traffic-based method places economies like Brunei, Suriname, and St. Kitts and Nevis at the top of its rankings, likely reflecting outlier behavior from a small number of power users or data sensitivities inherent in measuring traffic within very small populations. AI User Share, by contrast, focuses on the proportion of users in the entire working-age population using AI, helping to mitigate distortions caused by small traffic base or heavy individual usage. Our approach also intentionally excludes markets with insufficient data coverage or combines with similar, neighboring countries to create a regional average, offering a more stable and comparable signal of AI adoption across countries and regions. 

 To validate this expanded coverage and underscore the critical role of internet connectivity, we compared our telemetry-based AI User Share data to Google Trends data on search interest for "ChatGPT"~\supercite{noauthor_google_nodate}. Countries such as Sri Lanka, Peru, Nepal, Pakistan and Panama rank exceptionally high in relative interest in AI-related searches, corroborating our telemetry-based insight that Microsoft users in many developing nations show substantial enthusiasm for AI tools. This alignment strengthens the credibility of our AI User Share metric and underscores the latent demand for AI in less digitally inclusive regions. 

The substantial difference between country-level and connected-population AI usage suggests critical policy implications. Our findings imply that the major obstacle to AI adoption in many lower-income countries is not user demand but limited internet access. Consequently, policymakers and international organizations aiming to accelerate AI adoption should prioritize investments in digital infrastructure, internet affordability, and digital literacy. Improving internet infrastructure and affordability of digital access could unlock significant latent demand for AI technologies, fostering inclusive economic and social benefits. However, achieving this potential requires acknowledging that newly connected users may differ from early adopters in digital skills or access quality, potentially moderating the pace of AI uptake. Targeted investments should therefore encompass not only infrastructure and affordability but also digital literacy programs tailored to these future users.

Our analysis also highlights a strong correlation between AI adoption and economic strength, echoing previous findings ~\supercite{khan_artificial_2024, liu_who_2024}. Specifically, our results demonstrate a robust relationship (Spearman correlation = 0.83) between GDP per capita and AI user share, reinforcing the role of economic factors in influencing technological diffusion. However, our analysis reveals a pattern of more gradual increase in AI User Share at higher levels of income. Apart from notable outliers such as UAE and Singapore, most advanced economies cluster within the 25-45\% range. This pattern suggests that AI adoption may not continue to scale uniformly with economic development beyond a certain threshold. These observations motivate further investigation into the factors—such as regulatory environments, digital infrastructure, cultural factors, or education systems—that may shape the trajectory of AI engagement in advanced economies. 

Another distinctive advantage of our methodology is its near-real-time sensitivity. The rapid increase in China's AI usage immediately following DeepSeek's January 2025 launch illustrates this capability vividly. Unlike annual or semi-annual survey-based methods, our approach can capture market dynamics within weeks or even days of significant events. In theory, in regions with sufficiently dense data, trends could be monitored at even finer temporal resolutions—weekly or daily—enabling rapid assessment of policy impacts, product launches, and shifts in public sentiment toward AI. 

Nevertheless, our methodology carries some limitations. Because our metric originates with Microsoft telemetry, it is inherently biased toward desktop platforms and the Microsoft user demographic. Although we apply rigorous adjustments and scaling factors, our results implicitly assume that user behavior in Microsoft products approximates that in other platforms, which may not always hold true. Future iterations of this research could mitigate this limitation by integrating data from mobile app analytics providers such as Sensor Tower or leveraging web traffic analytics from tools like Semrush or SimilarWeb. Expanding data sources would help generalize our insights to the broader digital population and improve estimates of global AI usage comprehensively. 

In summary, our AI User Share metric provides a robust, real-time, usage-based measure of global AI adoption, complementing traditional survey-based methods and filling crucial measurement gaps. By highlighting both regional disparities and latent adoption potential among connected populations in developing economies, our findings offer policymakers actionable insights into prioritizing digital investments. Future work should leverage additional data sources to refine these estimates further, providing increasingly comprehensive and precise views of global AI diffusion. 


\printbibliography  

\clearpage  
\appendix  
\section*{Appendix}  
\addcontentsline{toc}{section}{Appendix}  
  
\section{List of Sites/Apps Included in Our Analysis}  

\begin{itemize}  
     \item Alice
    \item ChatGPT  
    \item Character.ai  
    \item Claude  
    \item ClOVA X  
    \item DeepSeek  
    \item ERNIE Bot (Yiyan.baidu)  
    \item GigaChat
    \item Google Gemini  
    \item Grok
    \item Khanmigo.ai  
    \item Meta.ai  
    \item Microsoft Copilot  
    \item Midjourney  
    \item Mistral.ai  
    \item NanoSemantics AI Assistant
    \item Perplexity  
    \item Tongyi Qianwen 
    \item Xiaowei  
\end{itemize}  

\section{Full Table of AI User Share}

  \begin{longtable}[l]{@{}
    >{\centering\arraybackslash}p{2.5cm}
    >{\centering\arraybackslash}p{3.8cm}  
    >{\centering\arraybackslash}p{1.6cm}
    >{\centering\arraybackslash}p{2.8cm}
    >{\centering\arraybackslash}p{2.0cm}@{}}
    \caption{AI User Share for all Countries 
} \\
    \toprule  
    \textbf{Economy} & \textbf{Region} & \textbf{AI User Share} & \textbf{AI User Share (Connected Population)} & \textbf{GDP Per Capita} \\
    \midrule  
    \endfirsthead  
    \midrule \multicolumn{5}{r}{Continued on next page} \\  
    \endfoot  
    \bottomrule  
    \endlastfoot  
Afghanistan\imputed & South Asia & 5.15\% & 30.75\% & \$416  \\
Albania & Europe \& Central Asia & 15.79\% & 18.84\% & \$8,575  \\
Algeria & Middle East \& North Africa & 11.25\% & 14.53\% & \$5,364  \\
Angola\imputed & Sub-Saharan Africa & 8.91\% & 19.49\% & \$2,308  \\
Argentina & Latin America \& Caribbean & 17.77\% & 19.71\% & \$14,187  \\
Armenia & Europe \& Central Asia & 6.22\% & 7.75\% & \$8,053  \\
Australia & East Asia \& Pacific & 34.50\% & 35.43\% & \$64,821  \\
Austria & Europe \& Central Asia & 29.15\% & 30.57\% & \$56,034  \\
Azerbaijan & Europe \& Central Asia & 14.23\% & 15.93\% & \$7,126  \\
Bangladesh & South Asia & 6.47\% & 14.30\% & \$2,551  \\
Belarus & Europe \& Central Asia & 7.56\% & 8.01\% & \$7,829  \\
Belgium & Europe \& Central Asia & 33.49\% & 34.81\% & \$54,701  \\
Benin\imputed & Sub-Saharan Africa & 8.71\% & 26.78\% & \$1,394  \\
Bolivia & Latin America \& Caribbean & 10.88\% & 15.31\% & \$3,686  \\
Bosnia And Herzegovina & Europe \& Central Asia & 18.24\% & 21.03\% & \$8,639  \\
Botswana & Sub-Saharan Africa & 12.83\% & 15.65\% & \$7,820  \\
Brazil & Latin America \& Caribbean & 15.55\% & 18.28\% & \$10,295  \\
Bulgaria & Europe \& Central Asia & 25.41\% & 30.40\% & \$15,886  \\
Burkina Faso\imputed & Sub-Saharan Africa & 8.71\% & 48.23\% & \$883  \\
Burundi\imputed & Sub-Saharan Africa & 6.37\% & 55.04\% & \$193  \\
Cambodia & East Asia \& Pacific & 4.60\% & 7.52\% & \$2,430  \\
Cameroon\imputed & Sub-Saharan Africa & 7.04\% & 17.38\% & \$1,737  \\
Canada & North America & 33.54\% & 35.46\% & \$53,431  \\
Central African Republic\imputed & Sub-Saharan Africa & 7.04\% & N/A & \$496  \\
Chad\imputed & Sub-Saharan Africa & 7.04\% & 51.21\% & \$681  \\
Chile & Latin America \& Caribbean & 19.57\% & 20.66\% & \$17,068  \\
China & East Asia \& Pacific & 15.40\% & 16.69\% & \$12,614  \\
Colombia & Latin America \& Caribbean & 20.41\% & 25.99\% & \$6,947  \\
Congo\imputed & Sub-Saharan Africa & 7.04\% & 18.92\% & \$2,478  \\
Congo (DRC)\imputed  & Sub-Saharan Africa & 7.04\% & 23.55\% & \$628  \\
Costa Rica & Latin America \& Caribbean & 25.11\% & 29.05\% & \$16,942  \\
Côte d’Ivoire & Sub-Saharan Africa & 10.84\% & 25.96\% & \$2,531  \\
Croatia & Europe \& Central Asia & 21.80\% & 25.76\% & \$21,865  \\
Cuba & Latin America \& Caribbean & 5.66\% & 7.90\% & N/A \\
Czechia & Europe \& Central Asia & 26.04\% & 29.38\% & \$31,591  \\
Denmark & Europe \& Central Asia & 26.65\% & 26.71\% & \$68,454  \\
Dominican Republic & Latin America \& Caribbean & 21.96\% & 24.00\% & \$10,718  \\
Ecuador & Latin America \& Caribbean & 16.99\% & 21.71\% & \$6,610  \\
Egypt & Middle East \& North Africa & 12.50\% & 17.04\% & \$3,457  \\
El Salvador & Latin America \& Caribbean & 14.59\% & 21.18\% & \$5,391  \\
Eritrea\imputed & Sub-Saharan Africa & 6.37\% & 32.19\% & N/A \\
Ethiopia\imputed & Sub-Saharan Africa & 6.37\% & N/A & \$1,272  \\
Finland & Europe \& Central Asia & 25.62\% & 27.26\% & \$52,926  \\
France & Europe \& Central Asia & 40.93\% & 45.48\% & \$44,691  \\
French Guiana\imputed & Latin America \& Caribbean & 8.34\% & N/A & N/A \\
Gabon & Sub-Saharan Africa & 12.31\% & 16.97\% & \$7,803  \\
Gambia & Sub-Saharan Africa & 10.64\% & 22.63\% & \$888  \\
Georgia & Europe \& Central Asia & 17.26\% & 20.86\% & \$8,284  \\
Germany & Europe \& Central Asia & 26.53\% & 28.23\% & \$54,343  \\
Ghana\imputed & Sub-Saharan Africa & 8.71\% & 12.69\% & \$2,260  \\
Greece & Europe \& Central Asia & 17.73\% & 20.40\% & \$23,401  \\
Guatemala & Latin America \& Caribbean & 13.69\% & 23.81\% & \$5,763  \\
Guinea\imputed & Sub-Saharan Africa & 8.71\% & 32.17\% & \$1,541  \\
Guinea-Bissau\imputed & Sub-Saharan Africa & 8.71\% & 26.58\% & \$951  \\
Guyana\imputed & Latin America \& Caribbean & 8.34\% & 10.19\% & \$20,765  \\
Haiti & Latin America \& Caribbean & 7.12\% & N/A & \$1,706  \\
Honduras & Latin America \& Caribbean & 12.37\% & 20.76\% & \$3,232  \\
Hungary & Europe \& Central Asia & 27.91\% & 29.60\% & \$22,142  \\
India & South Asia & 14.23\% & 20.13\% & \$2,481  \\
Indonesia & East Asia \& Pacific & 11.67\% & 15.87\% & \$4,876  \\
Iran & Middle East \& North Africa & 9.56\% & 11.94\% & \$4,466  \\
Iraq & Middle East \& North Africa & 10.29\% & 12.54\% & \$5,565  \\
Ireland & Europe \& Central Asia & 41.70\% & 43.03\% & \$103,888  \\
Israel & Middle East \& North Africa & 33.91\% & 37.99\% & \$52,642  \\
Italy & Europe \& Central Asia & 25.80\% & 28.66\% & \$39,003  \\
Jamaica & Latin America \& Caribbean & 22.24\% & 26.39\% & \$6,840  \\
Japan & East Asia \& Pacific & 16.74\% & 19.12\% & \$33,767  \\
Jordan & Middle East \& North Africa & 25.36\% & 27.31\% & \$4,456  \\
Kazakhstan & Europe \& Central Asia & 12.67\% & 13.54\% & \$12,919  \\
Kenya & Sub-Saharan Africa & 7.83\% & 21.65\% & \$1,952  \\
Kuwait & Middle East \& North Africa & 17.71\% & 17.76\% & \$33,730  \\
Kyrgyzstan & Europe \& Central Asia & 7.55\% & 8.52\% & \$1,970  \\
Laos & East Asia \& Pacific & 5.95\% & 9.29\% & \$2,067  \\
Lebanon & Middle East \& North Africa & 24.83\% & 29.40\% & N/A \\
Lesotho & Sub-Saharan Africa & 8.77\% & 17.89\% & \$916  \\
Liberia\imputed & Sub-Saharan Africa & 8.71\% & 35.97\% & \$772  \\
Libya & Middle East \& North Africa & 12.69\% & 14.30\% & \$6,173  \\
Lithuania & Europe \& Central Asia & 21.04\% & 23.60\% & \$27,786  \\
Madagascar\imputed & Sub-Saharan Africa & 8.91\% & 40.88\% & \$506  \\
Malawi\imputed & Sub-Saharan Africa & 8.91\% & 45.75\% & \$602  \\
Malaysia & East Asia \& Pacific & 18.29\% & 18.64\% & \$11,379  \\
Mali\imputed & Sub-Saharan Africa & 8.71\% & 24.69\% & \$869  \\
Mauritania\imputed & Sub-Saharan Africa & 8.71\% & 23.24\% & \$2,121  \\
Mexico & Latin America \& Caribbean & 16.72\% & 20.40\% & \$13,790  \\
Moldova & Europe \& Central Asia & 16.65\% & 20.52\% & \$6,729  \\
Mongolia & East Asia \& Pacific & 12.62\% & 15.10\% & \$5,839  \\
Morocco & Middle East \& North Africa & 10.55\% & 11.56\% & \$3,771  \\
Mozambique\imputed & Sub-Saharan Africa & 8.91\% & 41.86\% & \$623  \\
Myanmar & East Asia \& Pacific & 8.41\% & 14.23\% & \$1,233  \\
Namibia & Sub-Saharan Africa & 12.96\% & 19.76\% & \$4,168  \\
Nepal & South Asia & 12.27\% & 21.59\% & \$1,378  \\
Netherlands & Europe \& Central Asia & 36.33\% & 37.32\% & \$64,572  \\
New Zealand & East Asia \& Pacific & 37.57\% & 38.89\% & \$48,281  \\
Nicaragua & Latin America \& Caribbean & 9.95\% & 16.80\% & \$2,613  \\
Niger\imputed & Sub-Saharan Africa & 8.71\% & 36.42\% & \$643  \\
Nigeria\imputed & Sub-Saharan Africa & 8.71\% & 22.21\% & \$1,597  \\
Norway & Europe \& Central Asia & 45.34\% & 45.73\% & \$87,925  \\
Oman & Middle East \& North Africa & 22.60\% & 23.67\% & \$21,550  \\
Pakistan & South Asia & 9.65\% & 33.45\% & \$1,365  \\
Panama & Latin America \& Caribbean & 20.31\% & 25.66\% & \$18,686  \\
Papua New Guinea & East Asia \& Pacific & 7.15\% & 28.31\% & \$2,958  \\
Paraguay & Latin America \& Caribbean & 10.12\% & 12.33\% & \$6,276  \\
Peru & Latin America \& Caribbean & 13.42\% & 16.24\% & \$7,907  \\
Philippines & East Asia \& Pacific & 17.07\% & 20.22\% & \$3,805  \\
Poland & Europe \& Central Asia & 26.44\% & 29.56\% & \$22,057  \\
Portugal & Europe \& Central Asia & 22.44\% & 25.15\% & \$27,331  \\
Qatar & Middle East \& North Africa & 35.67\% & 35.79\% & \$80,196  \\
Romania & Europe \& Central Asia & 15.26\% & 16.65\% & \$18,404  \\
Russia & Europe \& Central Asia & 7.61\% & 8.06\% & \$13,817  \\
Rwanda & Sub-Saharan Africa & 6.05\% & 17.15\% & \$1,010  \\
Saudi Arabia & Middle East \& North Africa & 23.70\% & 23.70\% & \$32,094  \\
Senegal & Sub-Saharan Africa & 12.41\% & 20.20\% & \$1,706  \\
Serbia & Europe \& Central Asia & 19.69\% & 22.30\% & \$12,282  \\
Sierra Leone\imputed & Sub-Saharan Africa & 8.71\% & 40.53\% & \$758  \\
Singapore & East Asia \& Pacific & 58.63\% & 61.41\% & \$84,734  \\
Slovakia & Europe \& Central Asia & 22.07\% & 24.40\% & \$24,491  \\
Slovenia & Europe \& Central Asia & 24.55\% & 26.86\% & \$32,610  \\
Somalia\imputed & Sub-Saharan Africa & 6.37\% & N/A & \$597  \\
South Africa & Sub-Saharan Africa & 19.34\% & 25.27\% & \$6,023  \\
South Korea & East Asia \& Pacific & 25.93\% & 26.45\% & \$33,121  \\
South Sudan\imputed & Sub-Saharan Africa & 6.37\% & N/A & N/A \\
Spain & Europe \& Central Asia & 39.68\% & 41.22\% & \$33,509  \\
Sri Lanka & South Asia & 6.21\% & 11.95\% & \$3,828  \\
Sudan\imputed & Sub-Saharan Africa & 6.37\% & N/A & \$2,183  \\
Suriname\imputed & Latin America \& Caribbean & 8.34\% & 10.61\% & \$5,494  \\
Sweden & Europe \& Central Asia & 31.16\% & 32.48\% & \$55,517  \\
Switzerland & Europe \& Central Asia & 32.38\% & 33.18\% & \$99,565  \\
Syria & Middle East \& North Africa & 6.70\% & N/A & N/A \\
Taiwan & East Asia \& Pacific & 26.38\% & 27.21\% & \$32,442  \\
Tajikistan\imputed & Europe \& Central Asia & 5.15\% & 6.46\% & \$1,161  \\
Tanzania\imputed & Sub-Saharan Africa & 6.37\% & 22.62\% & \$1,224  \\
Thailand & East Asia \& Pacific & 9.12\% & 10.02\% & \$7,182  \\
Togo\imputed & Sub-Saharan Africa & 8.71\% & 23.46\% & \$986  \\
Tunisia & Middle East \& North Africa & 12.32\% & 16.82\% & \$3,978  \\
Türkiye & Europe \& Central Asia & 13.38\% & 15.26\% & \$13,106  \\
Turkmenistan\imputed & Europe \& Central Asia & 5.15\% & N/A & \$8,233  \\
Uganda\imputed & Sub-Saharan Africa & 6.37\% & 66.09\% & \$1,002  \\
Ukraine & Europe \& Central Asia & 9.14\% & 11.05\% & \$5,070  \\
United Arab Emirates & Middle East \& North Africa & 59.45\% & 59.45\% & \$49,041  \\
United Kingdom & Europe \& Central Asia & 36.38\% & 37.62\% & \$49,464  \\
United States & North America & 26.27\% & 28.04\% & \$82,769  \\
Uruguay & Latin America \& Caribbean & 20.91\% & 22.61\% & \$22,798  \\
Uzbekistan & Europe \& Central Asia & 5.73\% & 6.43\% & \$2,850  \\
Venezuela\imputed & Latin America \& Caribbean & 8.34\% & N/A & N/A \\
Vietnam & East Asia \& Pacific & 21.23\% & 24.99\% & \$4,282  \\
Zambia & Sub-Saharan Africa & 11.73\% & 34.10\% & \$1,331  \\
Zimbabwe & Sub-Saharan Africa & 6.95\% & 17.66\% & \$2,156  \\
        \bottomrule  
    \label{tab:ai_user_share_full}  
\end{longtable}  
\noindent\textsuperscript{\dag}\,Region-imputed estimate due to insufficient telemetry coverage.\\

\end{document}